\documentclass[12pt,titlepage,sort&compress]{article}
\usepackage{epsfig}

\def\ba{\begin{align}}
\def\ea{\end{align}}
\def\be{\begin{equation}}
\def\ee{\end{equation}}
\def\nn{\nonumber}
\def\bea{\begin{eqnarray}}
\def\eea{\end{eqnarray}}

\newcommand{\roughly}[1]{\mathrel{\raise.3ex\hbox{$#1$\kern-0.85em
  \lower1ex\hbox{$\sim$}}}}

\begin{document}

\title{Holomorphic and anti-holomorphic 
conductivity flows in the quantum Hall effect}

\author{{Brian P. Dolan}
\medskip
\and
\textit{\small Dept. of Mathematical Physics, National University of Ireland, Maynooth, Ireland.}\\
{\small and}\\
\textit{\small School of Theoretical Physics, 
Dublin Institute for Advanced Studies}\\
\textit{\small 10 Burlington Rd, Dublin, Ireland.}
}

\date{27th November 2010}
\maketitle

\begin{abstract} It is shown that the flow diagrams for the conductivities
in the quantum Hall effect, arising from two ostensibly
 very different proposals based on modular symmetry, are in fact identical.
The $\beta$-functions are different,  the rate at which the flow lines
are traversed are different, but the tangents to the flow lines are
the same in both cases, hence the flow diagrams are same in all aspects.

\bigskip

\noindent\hbox{Preprint no: DIAS-STP-10-12}

\medskip

\noindent\hbox{PACS nos: 73.43.Nq, 05.30.Fk, 05.30.Rt, 02.20.-a}

\end{abstract}

\section{Introduction}

The notion that the modular group, or a sub-group thereof, is relevant
to a description of the quantum Hall effect is now nearly twenty years
old, \cite{LutkenRossa,LutkenRossb,Lutken}.  The basic idea is that
a sub-modular group is an emergent symmetry that
maps between different phases of the 2-dimensional electron gas.
Denote the Hall conductivity in a homogeneous quantum Hall sample
by $\sigma_{xy}$ and the Ohmic conductivity by $\sigma_{xx}\ge 0$ 
(throughout this paper we use natural units for the conductivity,
with $\frac{e^2}{h}=1$).
The modular group acts on the complex
conductivity $\sigma=\sigma_{xy}+i\sigma_{xx}$ via
\be \label{gammaAction}\gamma(\sigma)=\frac{a\sigma +b}{c\sigma +d}\ee
for any four 
integers 
$a$, $b$, $c$ and $d$ satisfying $ad-bc=1$, these properties ensure
that $Im\bigl(\gamma(\sigma)\bigr)>0$ if $Im(\sigma)>0$.
Group multiplication is the same as matrix multiplication for 
$\gamma=\left(\begin{array}{cc} a & b \\ c & d \\ \end{array}\right)$,
though $\gamma$ and $-\gamma$ have the same effect in (\ref{gammaAction}).

The description so far allows for $\gamma$ in (\ref{gammaAction}) to be 
any element of the full modular group $\Gamma(1)\cong Sl(2,{\bf Z})/{\bf Z}_2$.  
However 
$\Gamma(1)$ is not the group relevant to the quantum Hall effect as 
a general element will not
in preserve the property of odd denominators for the
Hall conductivity at quantum Hall plateaux (exotic even denominator states
require special consideration in the modular setting \cite{ModularA}).  
The sub-group $\Gamma_0(2)\subset\Gamma(1)$, defined
by demanding that $c$ be even,
does preserve the parity of the denominators and
was identified as the correct group, at least for spin split samples, 
in \cite{LutkenRossb,Lutken}.
A key ingredient in the understanding of a sub-modular group 
as an emergent symmetry is
particle-vortex duality \cite{ModularA,ModularB,LRDuality}.

Modular symmetry has very important implications for the way in which
the conductivities change as the intrinsic scale of the microscopic
physics is changed, \cite{CAa}-\cite{LRProbes}.
It allows one to identify fixed points of the flow unambiguously
as any fixed point of $\Gamma_0(2)$ must be a fixed point of the flow.
By a fixed point of $\Gamma_0(2)$ we mean here a complex conductivity, 
$\sigma_*$, for which there exists an element $\gamma\in\Gamma_0(2)$ such
that $\gamma(\sigma_*)=\sigma_*$, for example $\sigma_*=\frac{n+i}{2}$ are fixed points for any integer $n$.  Such points are isolated in the upper-half
plane.  An example of the power of modular symmetry is the selection rule, that
transitions between two quantum Hall plateaux with filling fractions
$\nu=\frac p q$ and $\nu' = \frac{p'} {q'}$ is only allowed if 
$|p q' - q p'|=1$, \cite{SelectionRule}.
 
The first appearance of the modular group as an emergent symmetry 
in a two dimensional system was \cite{Cardy}, in the context
of a model chosen for properties similar to those expected
of QCD.  Some seven years later it was suggested
that the modular group, or a sub-group thereof,
should be a low energy emergent symmetry in
the quantum Hall effect \cite{WilczekShapere},
however the sub-group identified in \cite{WilczekShapere}
was not correct for a quantitative description of the quantum Hall effect.
A more detailed analysis was carried out in \cite{LutkenRossa}
and the particular sub-group of the modular group, $\Gamma_0(2)$ defined above, 
was identified as the correct sub-group in \cite{LutkenRossb,Lutken}.
At almost the same time as \cite{LutkenRossb} appeared, the 
``law of corresponding states'', based on effective field theory arguments, was put forward in \cite{KLZ} --- this generates $\Gamma_0(2)$ symmetry,
but complex notation and the language of modular symmetry was not used
in \cite{KLZ}. An alternative inhomogeneous action of $\Gamma(1)$
was given in \cite{FK}.
The relevance of other sub-groups of the modular group to the quantum Hall
effect was investigated in
\cite{GMW} and extensions to other systems (such as 2-d superconductors, 
\cite{ModularA,ModularB} quantum Hall bi-layers \cite{Bilayers} and graphene 
\cite{Graphene}.  The effects of electron spin and Zeeman splitting
were examined from the modular group point of view in \cite{Gammatwo,GammatwoB}.
A review of modular symmetry in the quantum Hall effect, and the relation
to $N=2$ supersymmetric Yang-Mills theory, is given in \cite{Review}.
  
A flow diagram for the quantum Hall effect, as the scattering length
associated with electron transport is varied, was conjectured in
\cite{Khmelnitskii}, but the normalisation of $\sigma_{xx}$ was not
determined. Scaling properties were further investigated theoretically
in \cite{PruiskenScaling} and experimentally in \cite{SUnivWTPP,Wanli}.
Macroscopically the electron scattering length, or in quantum
language the quantum coherence length of the electron wave function, 
can be controlled
by varying the temperature, at least until the point where the temperature
is so low that the coherence length becomes of the order of, or larger than,
the sample size. 

The first flow diagram compatible with $\Gamma_0(2)$ symmetry
appeared in \cite{LutkenRossb}.
The first quantitative 
investigations of the form of flow implied by sub-modular symmetry,
in \cite{CAa,CAb}, used gradient flow and the $c$-theorem for two-dimensional 
renormalisation group flow and required the introduction of
a metric on the upper-half conductivity plane.
An alternative suggestion used holomorphic $\beta$-functions to model 
the flow \cite{crossover,Gammatwo} and figure 1 is taken from \cite{Gammatwo}.  The flow presented in \cite{crossover} was compared with
experimental data in \cite{MurzinI}-\cite{Taiwan3} with encouraging results,
figures 2-4 are reproduced from \cite{MurzinI,MurzinF}. 
Gradient flow was revisited
in \cite{LRAntiHolomorphic,LRGeometricScaling,LRImplications} 
using an anti-holomorphic potential
and the
resulting flow diagram compared with the experimental data 
\cite{MurzinI,MurzinF} in \cite{LRImplications}, with equally good agreement,
see figures 5 and 6 reproduced from \cite{LRImplications}.
Indeed the similarity between the flow diagrams in \cite{crossover,Gammatwo}
and \cite{LRImplications} is remarkable, but this should perhaps not be so 
surprising as they
both rely on the same underlying symmetry, $\Gamma_0(2)$, which 
is very restrictive.  In this paper it will be shown that in fact
these two flow diagrams are identical, despite the fact that the underlying
$\beta$-functions are different.  The only difference in the integrated flow
is the {\it rate} at which the flow lines are traversed as the length scale is
changed, the flow diagram itself is identical in both cases.

\section{$\beta$-functions}

Any $\beta$-function 
$$\beta(\sigma,\bar\sigma)=\frac{d\sigma}{ds}$$ 
compatible with $\Gamma_0(2)$ must transform as
\be \label{betatransform}
\frac{d\gamma(\sigma)}{ds}=
\beta\bigl(\gamma(\sigma),\gamma(\bar\sigma)\bigr)
={1\over (c\sigma +d)^2}\beta(\sigma,\bar\sigma),\ee
where $\gamma(\sigma)=\frac{a\sigma +b}{c\sigma +d}$ is a $\Gamma_0(2)$
transformation. The real parameter $s$ here is the logarithm of some length
associated with the underlying physics, such as the electron scattering
length (a function of temperature). 
$\beta$-functions compatible with $\Gamma_0(2)$ symmetry were first
discussed in \cite{CAa}, in the context of gradient flow.
Some general properties of $\beta$-functions satisfying (\ref{betatransform}),
including the semi-circle law, were derived in \cite{semicircle}.

The function 
\[
\label{fsigma}
f(\sigma)=-{\vartheta_3^4\vartheta_4^4\over\vartheta_2^8}=
-{1\over 256 q^2}\prod_{n=1}^\infty{1\over \bigl(1+q^{2n}\bigr)^{24}}
\]
is invariant under $\Gamma_0(2)$, \cite{Rankin} (definitions and relevant properties
of Jacobi $\vartheta$-functions are summarised in the appendix).
Since $d\bigl(\gamma(\sigma)\bigr)\rightarrow \frac{d\sigma}{(c\sigma +d)^2}$
under $\Gamma_0(2)$ transformation
it is immediate that $f'$ must transform as
\be 
\frac{df}{d\sigma } \quad\longrightarrow\quad (c\sigma +d)^2 \frac{df}{d\sigma },\ee
{\it i.e.} it is a modular function of weight $2$.
The first use of such modular functions in the
context of (\ref{betatransform}) was \cite{CAa}
where the function
\be \label{Weighttwo} 
{\cal E}(\sigma):=\frac{1}{2\pi i}\frac{f'}{f}=1+24\sum_{n=1}^\infty \frac{n q^{2n}}{(1+ q^{2n})}\ee
was considered.

A second function, with the same transformation properties as $f'$
but which is not analytic, was also
considered in \cite{CAa} 
\be \label{Hecke}
{\cal H}(\sigma,\overline\sigma):=\frac{1}{\pi Im(\sigma)} + 16\sum_{n=1}^\infty\frac{nq^{2n}}{1-q^{4n}}
= \frac {i}{\pi}\frac{d}{d\sigma}\ln \Bigl((\sigma-\overline\sigma)^2\vartheta_3^2 \vartheta_4^2\Bigr).\ee

All attempts at constructing $\beta$-functions compatible with $\Gamma_0(2)$
symmetry to date have focused on (\ref{Weighttwo}) and (\ref{Hecke}).

\subsection{Holomorphic $\beta$-function}
\medskip
A $\beta$-function compatible with (\ref{betatransform})
is the holomorphic form of weight -2
\be \label{holomorphicbeta}
\widetilde\beta(\sigma)=-\frac{f}{f'}=\frac{1}{i\pi}
{1\over (\vartheta_3^4 + \vartheta_4^4)},\ee
where $f'={d f\over d\sigma}$, \cite{crossover,Gammatwo}. 

Equation (\ref{holomorphicbeta}) can immediately be integrated to give
\be
\frac{f'}{f}d\sigma=\frac{df}{f}=-ds\qquad\Rightarrow
\qquad f(\sigma)=Ce^{-s},\ee
where $C$ is a (complex) constant. The integral curves of the flow
are curves on which the complex phase of $f(\sigma)$ is constant.
These are easily plotted as a contour plot of the phase of $f(\sigma)$,
\cite{crossover,Gammatwo}, and the relevant part of figure 2 from 
\cite{Gammatwo} is shown in
figure 1.

This holomorphic flow was compared with experimental data 
for temperature flow in the integral quantum effect in \cite{MurzinI} and the fractional effect in \cite{MurzinF}: figure 2 is taken from \cite{MurzinI} and figures 3 and 4 are from \cite{MurzinF}.  For a short review see
\cite{Mini-review}.

\bigskip
\subsection{ Anti-holomorphic gradient flow}
\medskip
Following on from \cite{CAa} non-holomorphic gradient flow $\beta$-functions, 
compatible with $\Gamma_0(2)$ symmetry, were
further explored in \cite{CAb}
and a specific form was proposed in
\cite{LRAntiHolomorphic}
and further investigated in \cite{LRGeometricScaling,LRImplications,LRProbes}. 
The $\beta$-function proposed in \cite{LRAntiHolomorphic}
was integrated numerically and plotted in \cite{LRImplications}.

This $\beta$-function is obtained by first considering the holomorphic
function for $\Gamma_0(2)$ of weight +2 in (\ref{Weighttwo}), 
${\cal E}(\sigma)$,
which satisfies
\[\label{weighttwo}
{\cal E}\bigl(\gamma(\sigma)\bigr)=(c\sigma +d)^2 {\cal E}(\sigma).\]
This is then used to generate anti-holomorphic gradient flow
\be\label{gradientbeta}
\beta^\sigma(\sigma,\overline\sigma) = -i\, G^{\sigma\bar\sigma}\overline{{\cal E}(\sigma)}
=-{1\over 2\pi}G^{\sigma\bar\sigma}\overline{(\partial_\sigma \ln f)},\ee
where $G_{\sigma\bar\sigma}$ is a metric on the $\sigma$-plane.

A natural choice of metric is the modular invariant line element
\be
{d\sigma d\overline\sigma \over \bigl(Im\,{\sigma}\bigr)^2}
=G_{\sigma\overline\sigma}d\sigma d\overline\sigma \qquad\Rightarrow\qquad G^{\sigma\bar\sigma}= (Im\,\sigma)^2,\ee
but the choice of metric does not actually affect the flow diagrams, as we shall
see.  Under modular transformations
\be
(Im\, \sigma)^2 \quad \longrightarrow \quad {(Im\,\sigma)^2\over |c\sigma +d|^4},\ee
so, since 
\be
\overline{{\cal E}\bigl(\gamma(\sigma)\bigr)}=
(c\overline\sigma +d)^2 \overline{{\cal E}(\sigma)},\ee
(\ref{gradientbeta}) transforms correctly as
\be
\beta^\sigma\bigl(\gamma(\sigma),\overline{\gamma(\sigma)}\bigr)=
\frac{1}{ (c\sigma+d)^2}
\beta^\sigma(\sigma,\overline\sigma).\ee

 Equation (\ref{gradientbeta}) was integrated numerically in \cite{LRImplications} and the resulting flow
is plotted, together with the experimental data from \cite{MurzinI} and \cite{MurzinF},
in figures 5 and 6.  

\subsection{Comparison of $\beta$-functions}

The similarity with the flow plotted from (\ref{holomorphicbeta}) in 
figure 1 and from (\ref{gradientbeta}) in figures 5 and 6 is clear and we now
show that these flows are identical, even though the two underlying 
$\beta$-functions differ.  The crucial point is that the shape of the
flow lines depends only on their tangents at any point, not on the
individual values of $\dot\sigma_{xy}=Re\,\beta$ and
$\dot\sigma_{xx}=Im\,\beta$ separately, but only on the ratio
\be
{\dot\sigma_{xx}\over \dot\sigma_{xy}}:={Im\,\beta\over Re\,\beta}\ee
and this ratio is the same for both (\ref{holomorphicbeta}) and (\ref{gradientbeta}).  To see this first
note that $G^{\sigma\overline\sigma}$, being real, drops out in the ratio 
for (\ref{gradientbeta}) so we only need consider 
$\frac{Im\bigl(-\overline{f'/f}\bigr)}{ Re\bigl(-\overline{f'/f}\bigr)}$
for (\ref{gradientbeta}) and compare this with $\frac{Im(-f/f')}{ Re(-f/f')}$ for (\ref{holomorphicbeta}).
But for any complex number $w$ 
\be
{Im(w)\over Re(w)}=
{Im\left({{1\over \overline w}}\right)\over Re\left({{1\over \overline w}}\right)}\ee
and hence the ratio is the same for both (\ref{holomorphicbeta}) and (\ref{gradientbeta}).
The tangent to the resulting flow lines is therefore the same at
every point in the upper-half $\sigma$-plane and so the plots of
the two flows are necessarily identical.
Of course integrating the equations gives different solutions, but the
solutions only differ in the rate at which the flow lines are traversed,
not in the shape of the plots.

The nature of the repulsive fixed points at $\sigma_*=\frac{1+i}{2}$, 
and its images
under $\Gamma_0(2)$, can be investigated in detail by using the approximate form
\[ f(\sigma_* + \epsilon)=\frac 1 4 - a \epsilon^2 + o(\epsilon^4),\]
with $a=\frac{\left\{\Gamma\left(\frac 1 4 \right)\right\}^8}{64\pi^4}$
positive, \cite{crossover,Taniguchi}.
Then $-\frac{f}{f'}\approx \frac 1 {8a\epsilon} = 
\frac{\overline\epsilon}{8 a |\varepsilon|^2}$ giving the flow lines of
a hyperbolic fixed point as shown in figure 7.  
Of course one gets the same
form by using gradient flow with $-\frac{\overline f'}{\overline f}\approx
8a\overline\epsilon$, they differ by an overall real factor but
the geometry of the flow is the same. The fact that the former case has
a divergent factor $\sim\frac 1 {|\epsilon|^2}$ is not a pathology --- at
zero temperature one expects a discrete jump from $\sigma_{xy}=0$
to $\sigma_{xy}=1$ as the magnetic field is varied \cite{crossover}.

Near the origin $\sigma\approx\epsilon$ one has 
$f\approx -16e^{-\frac{i\pi}{\sigma}}$, \cite{crossover}, and
$\frac{f'}{f}= \frac{i\pi}{\sigma^2}$ so
the holomorphic $\beta$-function gives 
\[ \widetilde\beta\approx -\frac{f}{f'}=\frac{i\sigma^2}{\pi} \]
while anti-holomorphic gradient flow gives
\[ \beta\approx \Bigl(Im(\sigma)\Bigr)^2 \frac{i\sigma^2}{2|\sigma|^4}.\]
The later is invariant under a constant (real) rescaling of $\sigma$, while
the former goes to zero as $\sigma\rightarrow 0$.  

\section{Conclusion}

In conclusion it has been shown that the geometry of the two different
conductivity flows presented in \cite{crossover} and \cite{LRAntiHolomorphic}
are identical, despite the different $\beta$-functions.
The only differences lie in the rate at which the flow lines
are traversed, not in their shape.  

It is a pleasure to acknowledge many useful and illuminating 
discussions about the quantum Hall effect with Cliff Burgess.

\appendix

\section{Appendix}

Jacobi $\vartheta$-functions are defined as
\bea
\vartheta_2&=&2\sum_{n=0}^\infty q^{(n+\frac{1}{2})^2}=
2q^{1\over 4}\prod_{n=1}^\infty\bigl(1-q^{2n}\bigr)\bigl(1+q^{2n}\bigr)^2,\\
\vartheta_3&=&\sum_{n=-\infty}^\infty q^{n^2}=
\prod_{n=1}^\infty\bigl(1-q^{2n}\bigr)\bigl(1+q^{2n-1}\bigr)^2,\\
\vartheta_4&=&\sum_{n=-\infty}^\infty (-1)^nq^{n^2}=
\prod_{n=1}^\infty\bigl(1-q^{2n}\bigr)\bigl(1-q^{2n-1}\bigr)^2, 
\eea
with $q:=e^{i\pi\sigma}$ 
(the conventions are those of \cite{WW}, except that $\tau$
there is replaced by $\sigma$ here).

The $\vartheta$-functions satisfy the relation
\be\vartheta_3^4=\vartheta_2^4+\vartheta_4^4\ee
and have the following transformations under 
$T:\sigma\rightarrow \sigma+1$ and $S:\sigma\rightarrow -\frac{1}{\sigma}$
\bea
\vartheta_2(\sigma+1)=e^{i\pi\over 4}\vartheta_2(\sigma) ,
&\qquad & \vartheta_2\left(-{1\over\sigma}\right)=\sqrt{-i\sigma}\;\vartheta_4(\sigma),\\
\vartheta_3(\sigma+1)=\vartheta_4(\sigma),
\;\quad&\qquad & \vartheta_3\left(-{1\over\sigma}\right)=\sqrt{-i\sigma}\;\vartheta_3(\sigma),\\
\vartheta_4(\sigma+1)=\vartheta_3(\sigma), 
\;\quad&\qquad & \vartheta_4\left(-{1\over\sigma}\right)=\sqrt{-i\sigma}\;\vartheta_2(\sigma). \eea

Using these properties it is not difficult to show that the function
\bea
f(\sigma)=-\frac{\vartheta_3^4\vartheta_4^4}{\vartheta_2^8}&=&
-\frac{1}{256 q^2}\prod_{n=1}^\infty\frac{(1-q^{4n-2})^8}{\bigl(1+q^{2n}\bigr)^{16}}\nn \\
&=&-\frac{1}{256 q^2}\prod_{n=1}^\infty\frac{(1-q^{2n})^8}{\bigl(1+q^{2n}\bigr)^{16}(1-q^{4n})^8}\nn \\
&=&-\frac{1}{256 q^2}\prod_{n=1}^\infty{1\over \bigl(1+q^{2n}\bigr)^{24}}\nn
\eea
is invariant under $T$ and $ST^2S$, $\sigma\rightarrow {1\over 1-2\sigma}$.

\vfill\eject

\ 

\includegraphics{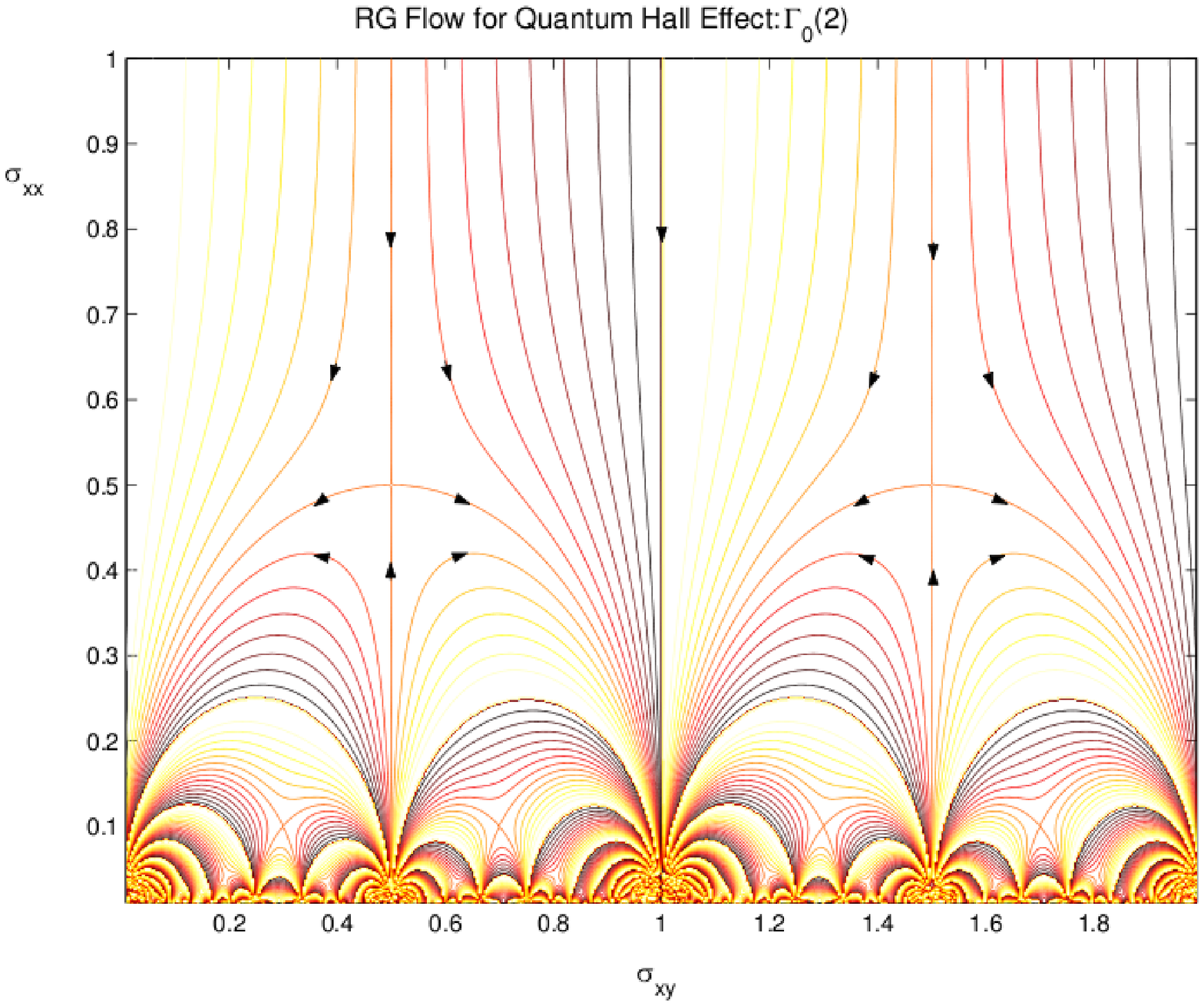}

\vskip 13cm

\centerline{\bf Figure 1.}

\vfill\eject

\ 
\includegraphics{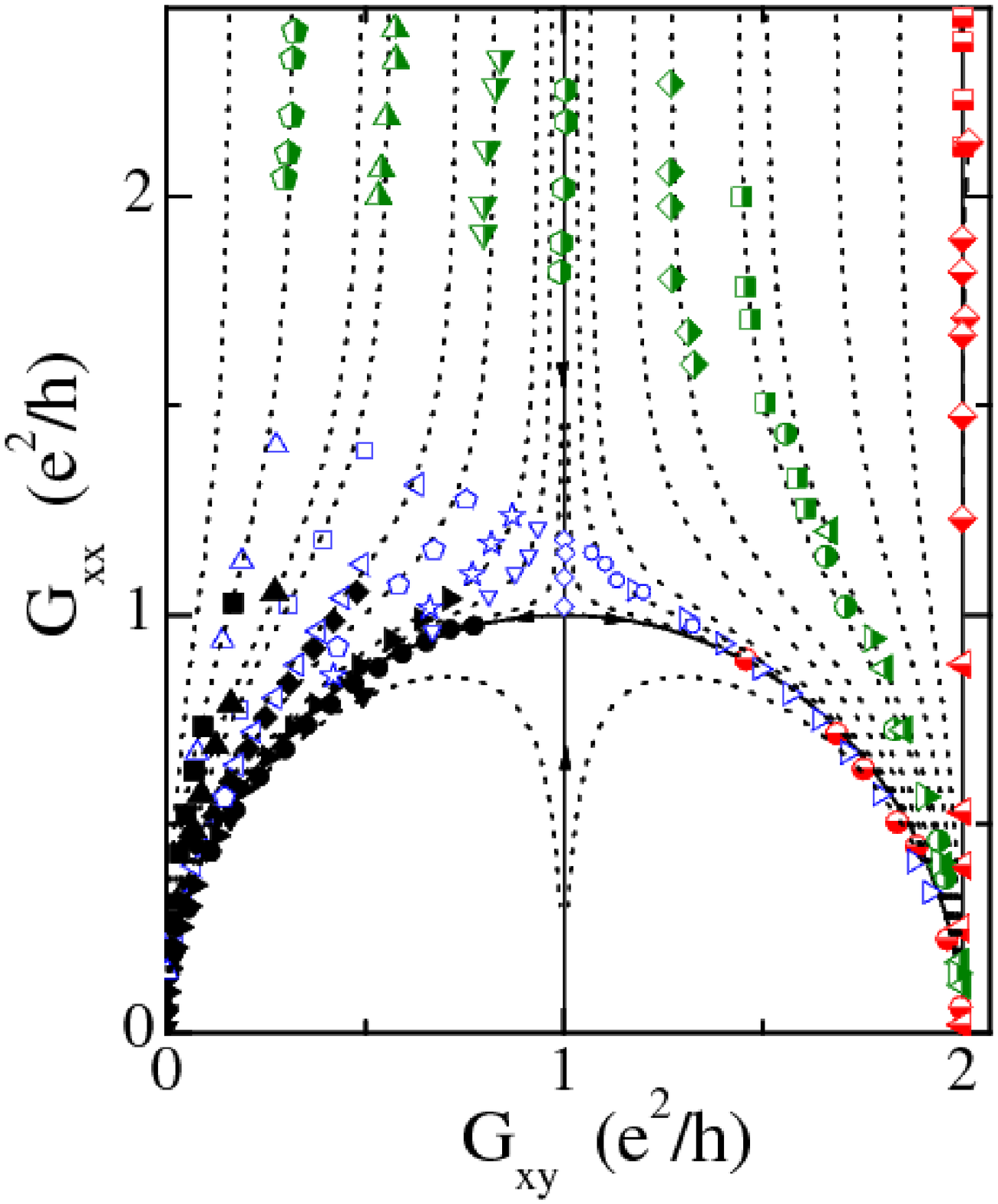}

\vskip 16cm 
\centerline{\bf Figure 2.}

\vfill\eject

\ 
\includegraphics{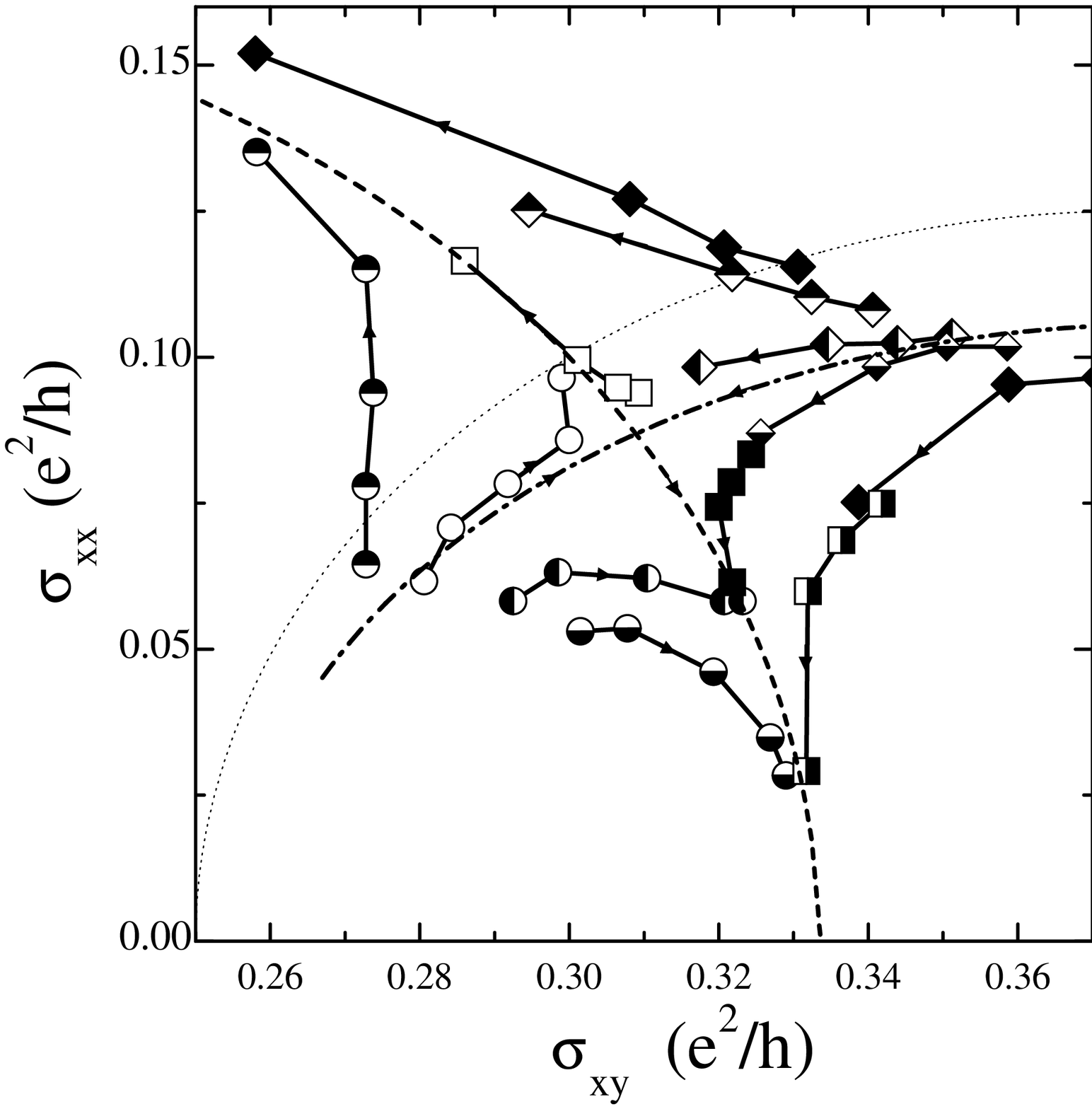}

\vskip 17cm
\centerline{\bf Figure 3.}
\vfill\eject

\ 
\includegraphics{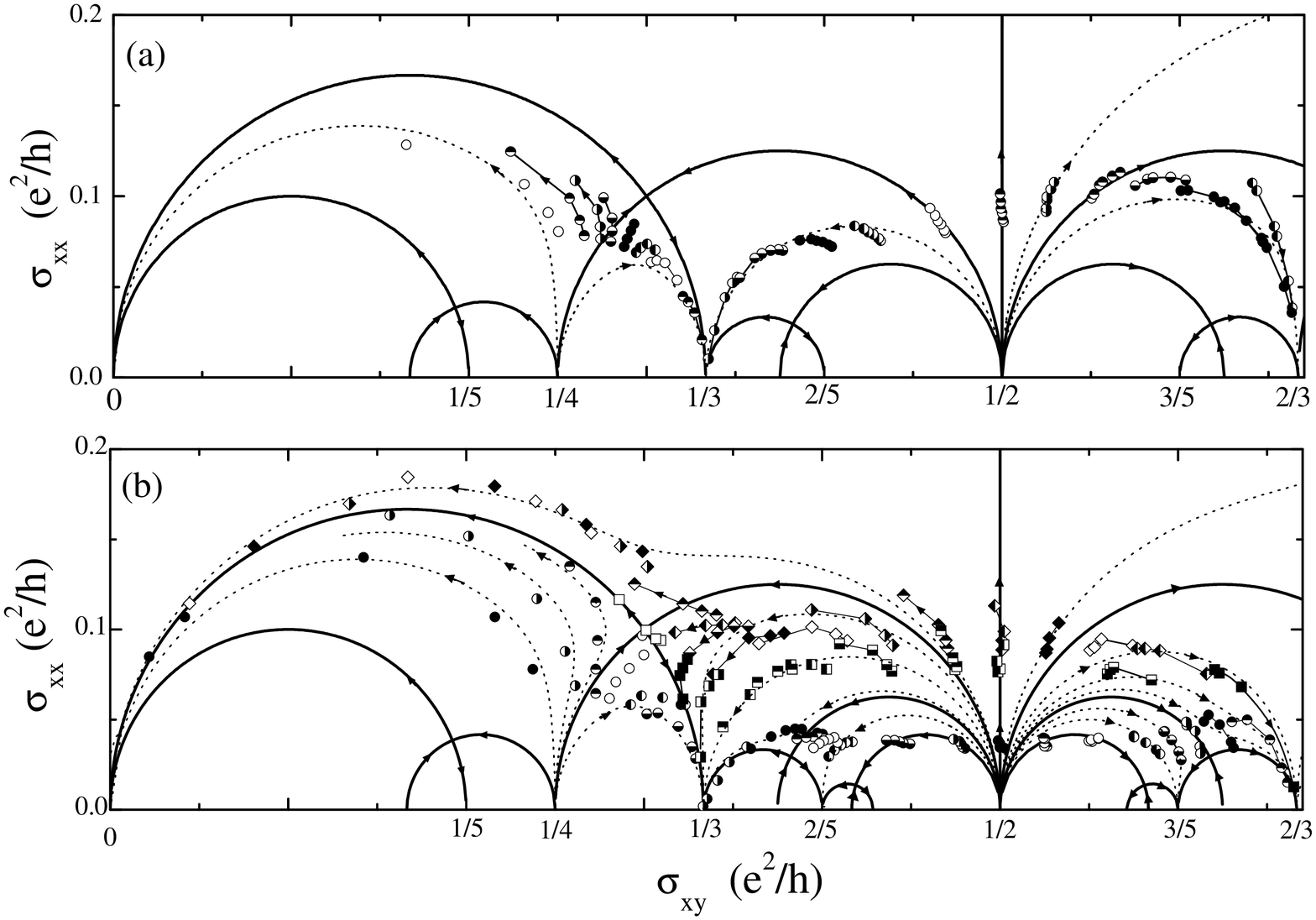}

\vskip 18cm
\centerline{\bf Figure 4.}

\vfill\eject

\ 
\includegraphics{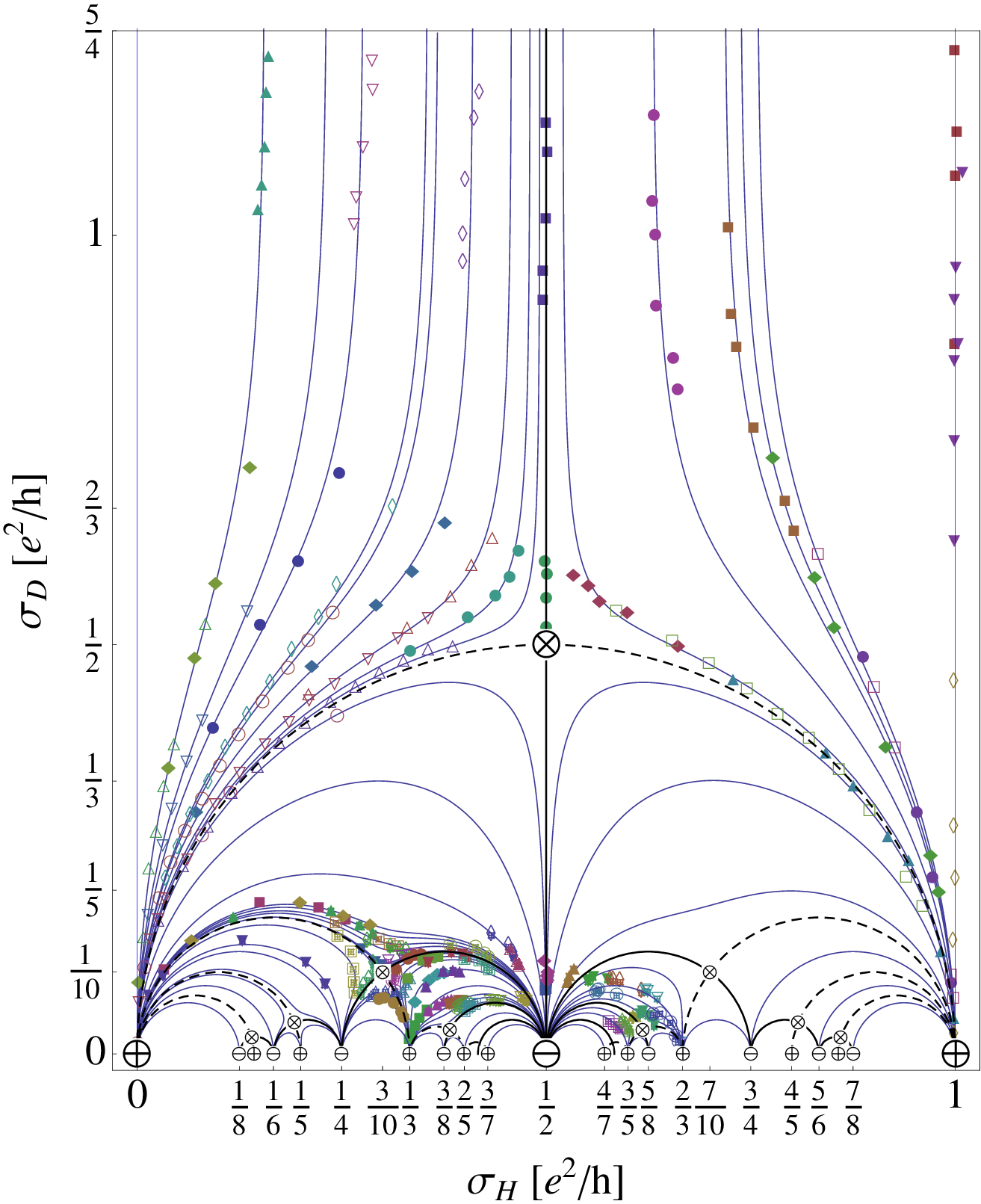}

\vskip 17cm
\centerline{\bf Figure 5.}

\vfill\eject

\ 

\includegraphics{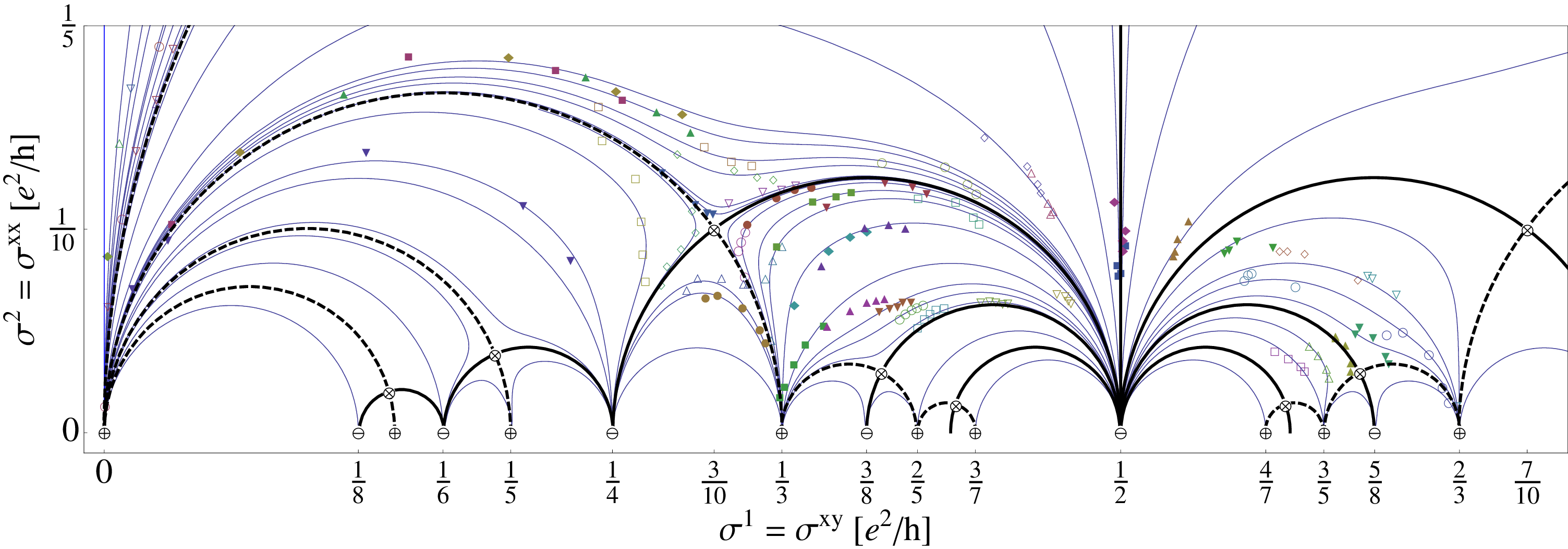}

\vskip 10cm

\centerline{\bf Figure 6.}

\hfill\eject

\ 

\includegraphics{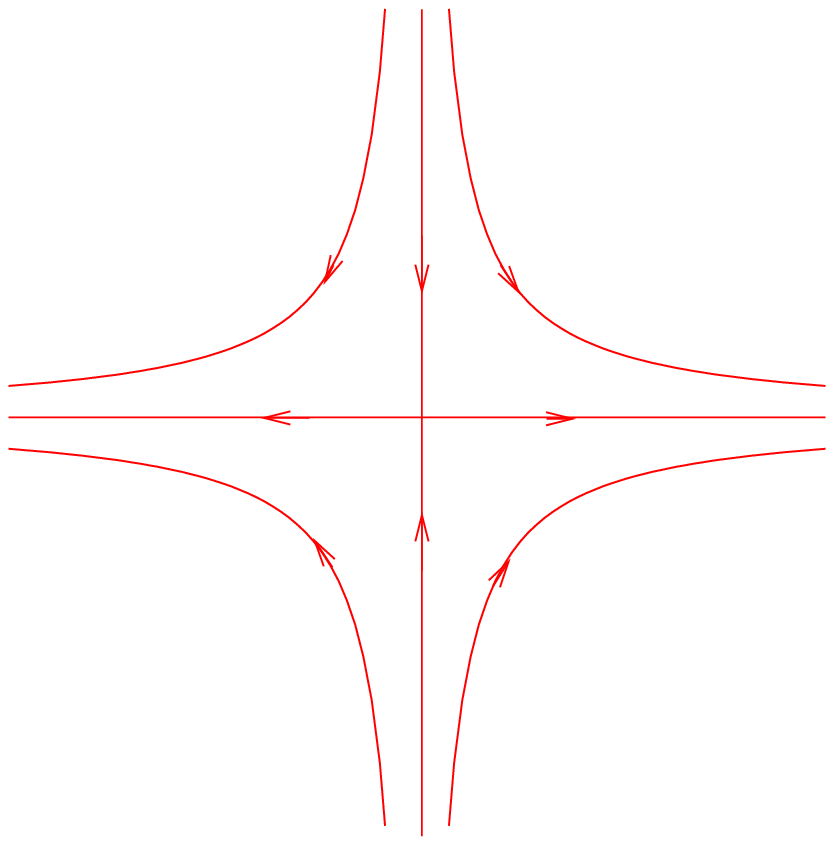}

\vskip 13cm

\centerline{\bf Figure 7.}

\end{document}